# Should children follow their parents' research paths? Intergenerational research continuity and divergence in academic families

Er-Te Zheng[1], Xiaorui Jiang[1], Zhichao Fang[2,3*], Mike Thelwall[1*]

[1] School of Information, Journalism and Communication, University of Sheffield, Sheffield, UK.

[2] School of Information Resource Management, Renmin University of China, Beijing, China.

[3] Centre for Science and Technology Studies (CWTS), Leiden University, Leiden, The Netherlands.

Corresponding authors: Zhichao Fang, E-mail: z.fang@cwts.leidenuniv.nl; Mike Thelwall, E-mail: m.a.thelwall@sheffield.ac.uk

## Abstract

How academic advantages are transmitted within families is usually studied as occupational inheritance, but it is not clear whether scholarly research orientations persist across generations and if it is an advantage when it does. To address this, we link Wikidata kinship records with OpenAlex bibliometric profiles to study 3,229 documented parent–child scholar pairs and 488,659 publications. Field-level research similarity was evident but not universal: whilst the median similarity was 0.546, 25.3% of parent–child pairs had no Field overlap (i.e., similarity 0). These pairs were substantially more similar than publication-period-matched comparison pairs (median 0.098). Direct academic interaction was uncommon: 10.4% of parent–child pairs had co-authored, 9.8% of children had cited their parents, and 6.9% of parents had cited their children. Nevertheless, each 0.1 increase in Field similarity was associated with 38–39% higher adjusted odds of co-authorship and cross-citation. There was also intergenerational continuity in academic achievement and recognition. Parents' publication volume and field-normalized citation impact were positively associated with those of their children. Children of national academy members had approximately twice the odds of becoming national academy members themselves (Odds Ratio = 2.04), while children of prizewinning parents had 46% higher odds of winning prizes (Odds Ratio = 1.46). However, children of national academy members showed lower research similarity to their parents. Greater research differentiation was associated with higher field-normalized citation impact among children, but not with publication output or higher odds of academic recognition. Academic families therefore appear to transmit resources and advantages with the sole exception that diverging from parental fields seems to confer a citation advantage.

## Keywords

Academic families; Parent–child scholars; Intergenerational research similarity; Science of science; Sociology of science; Scientometrics

## Introduction

Intergenerational occupational transmission is a central concern in family sociology. Previous studies have shown that children are more likely to enter occupations with socioeconomic status similar to that of their parents, while those from professional families may face fewer barriers to entering professional careers (1–3). Such intergenerational continuity is especially evident in highly specialized professions, including law, medicine, and academia, and has been observed across different social and cultural contexts (4–8). Family resources, including cultural and social capital, may facilitate children's entry into and early advancement within these professions (3, 9, 10).

Academic families provide a distinctive setting in which inherited advantages may either reinforce intellectual continuity or facilitate the pursuit of independent research paths. According to cumulative advantage theory (11), the academic authority, professional networks, and institutional resources accumulated by scholars provide their children with a substantial initial advantage if they choose to join academia. Because these resources are often embedded within particular disciplinary communities, the children of scholars may benefit most from pursuing research related to that of their parents, perhaps encouraged by parents for this reason. In contrast, niche partitioning theory (12), which suggests that children might try to find different niches to avoid successful parents, and the scientific community's emphasis on originality and independence (13), both predict a degree of intergenerational differentiation. Thus, following a parent scholar's established research path may expose children to direct competition, constrain the development of an independent academic identity, and leave them overshadowed by parental prestige. These perspectives generate competing expectations regarding whether academic families reproduce established research paths or promote intergenerational differentiation. Which pattern predominates, and the degree of differentiation, remain open empirical questions.

Large-scale bibliometric research on intergenerational academic relationships has primarily examined mentor–mentee pairings. Existing studies show that mentors' academic influence and resources are associated with their mentees' subsequent research performance and career outcomes (14–16). Research similarity between mentors and mentees may also shape these outcomes. Although intellectual proximity can facilitate the transmission of knowledge and expertise, and has been associated with mentees' early-career funding success (17), developing research that is distinct from that of one's mentor has been associated with a greater likelihood of attaining academic recognition, such as prizewinning (18). These findings suggest that academic successors may benefit from access to intellectual and social resources while also needing to establish independent research identities.

Mentor–mentee relationships, however, capture professional socialization primarily during and after graduate education. Parent–child relationships may constitute an additional channel of academic socialization that begins earlier and extends beyond formal training. Children raised in academic families may be exposed to disciplinary knowledge, research practices, professional networks, and understandings of academic careers before entering higher education. Such family environments may contribute to the formation of research preferences and academic habitus (19, 20), as well as providing access to tacit knowledge that is difficult to acquire through formal instruction alone (21).

Research on intergenerational transmission has largely focused on occupational inheritance, including whether children of scholars enter academia, rather than any tendency to follow similar research paths. Little is known about whether scholarly parents and children tend to pursue similar research paths and whether intergenerational research similarity is associated with the academic achievement of scholars' children. This gap partly reflects the lack of large-scale datasets linking verified parent–child relationships to comprehensive publication oeuvres, as well as suitable methods for comparing research orientations across generations. Addressing these questions shifts attention from the reproduction of academic occupations to the continuity and divergence of research interests within academic families.

This study addresses these gaps by linking parent–child relationships from Wikidata to OpenAlex author profiles and publication records. We assess research similarity using OpenAlex's hierarchical classification of Domains, Fields, Subfields, and Topics. Drawing on the linked data, we ask three questions: How similar are the research fields of parent and child scholars? How is this similarity associated with co-authorship and cross-citation? How do parents' academic achievements and parent–child research similarity relate to children's academic performance and recognition?

# Results

## Parent–child scholars are more similar than matched comparison pairs

Field similarity was available for 3,161 of the 3,229 parent–child pairs and varied substantially. The median similarity was 0.546; 52.6% of pairs had similarity above 0.5, while 25.3% had no Field overlap (Fig. 1A). Similarity also varied by parental Domain, with the highest median similarity among Social Sciences (Fig. 1B). The Social Sciences had the highest same-Domain "retention", with 78.3% of children remaining in their parents' Domain, but substantial intergenerational cross-Domain movement was present in all Domains (Fig. 1C).

To test whether this pattern exceeded what would be expected among scholars from comparable historical periods, we compared each actual pair with publication-decade matched random comparison pairs. Actual parent–child pairs were substantially more similar than the matched baseline (median 0.546 versus 0.098; $p = 0.001$). The result remained under stricter baselines matching additionally on children's gender and continent (Fig. S3 and Table S7). Field similarity had only a weak negative association with children's mean publication year ($r = -0.042$; Figs. 1D and S4D), and gender composition did not clearly differentiate similarity (Fig. S9).

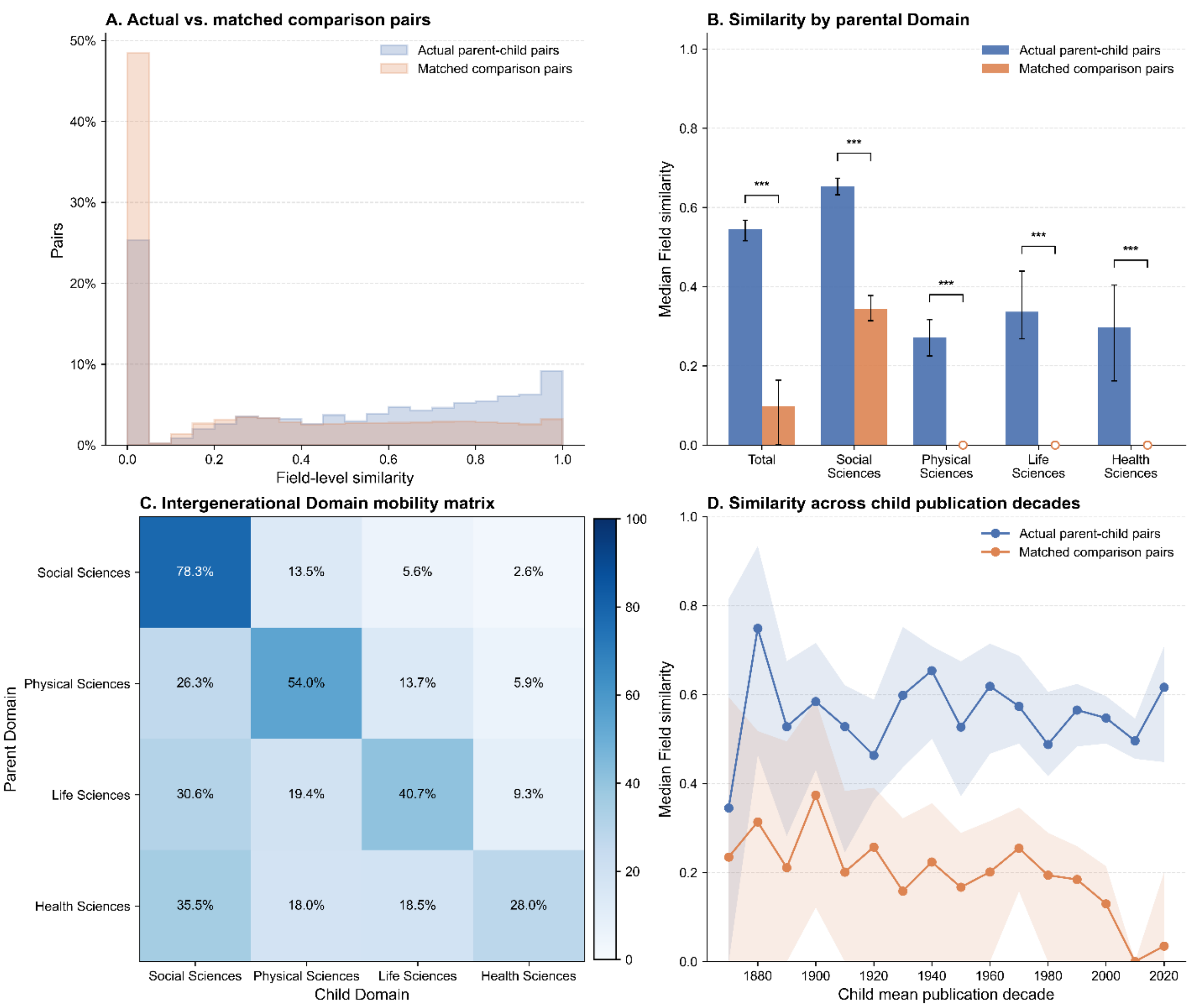

**Figure 1.** Intergenerational research similarity between parent and child scholars. (A) Distribution of observed parent–child Field similarity compared with publication-period matched random comparison pairs. (B) Median Field similarity by parental Domain for actual parent–child pairs and the matched baseline. (C) Intergenerational Domain mobility matrix showing row percentages from parent to child Domain. (D) Median Field similarity across children's publication decades. For panels B and D, intervals for actual parent–child pairs are 95% bootstrap confidence intervals for the median, whereas intervals for matched comparison pairs represent the 2.5th–97.5th percentiles of median similarity across 1,000 random-matching iterations.

## Research similarity is associated with academic interaction

Direct academic interaction between parents and children was uncommon but strongly patterned by research proximity. Overall, 10.4% of pairs had co-authored at least one publication (Fig. 2A). Children cited their parents' work in 9.8% of pairs, while parents cited their children's work in 6.9%; this directional difference was significant (Fig. 2C). Interaction was lowest among pairs in which the parent was in the Social Sciences, despite the Social Sciences showing the highest intergenerational retention (Fig. 2 A and C).

Parent–child pairs with higher Field similarity (≥0.5) had higher co-authorship rates than lower-similarity pairs (15.8% versus 4.7%; Fig. 2B). The same pattern appeared for parent-to-child citation (10.5% versus 3.3%) and child-to-parent citation (14.4% versus 5.2%; Fig. 2D). In multivariable models adjusting for both generations' Domains, gender, mean publication years, career spans, publication counts, and duration of overlap between the parents' and children's publication-career intervals, a 0.1 increase in Field similarity was associated with 38-39% higher odds of co-authorship and cross-citation (Table S3).

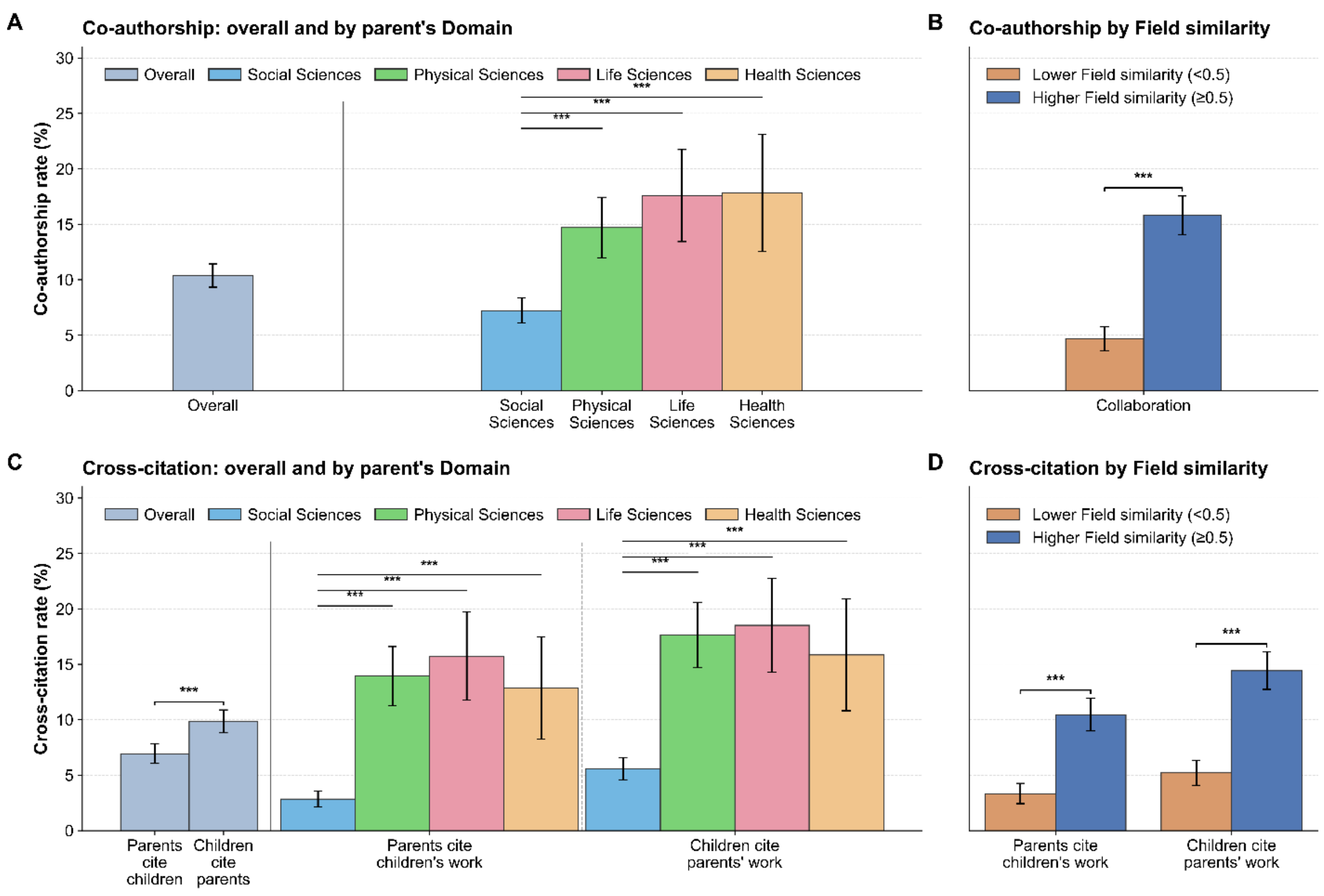

**Figure 2.** Co-authorship and cross-citation between parent and child scholars across parental Domains and Field-similarity groups.

## Academic achievement and recognition show continuity, while research similarity has selective associations with children's academic outcomes

Parents' and children's bibliometric performance was positively associated (Fig. 3 A–B). The correlation was modest for publication count ($r = 0.199$) and stronger for mean field-weighted citation impact (FWCI; $r = 0.321$). These associations remained significant after adjustment for publication period, career span, gender, Domain, and parental recognition (Table S4). Academic recognition had a similar pattern of intergenerational continuity: children of national academy members were more likely to become national academy members themselves, and children of prizewinning parents were more likely to receive recorded prizes (Fig. 3 C–D). In controlled models, these associations were specific to the corresponding form of recognition. Children of national academy members had approximately twice the odds of national academy membership compared with children of non-members (OR = 2.04), children of prizewinning parents had 46% higher odds of prizewinning status than children of non-prizewinning parents (OR = 1.46), while cross-form associations were not statistically significant (Table S5).

Field similarity had selective associations with children's academic outcomes (Fig. 4). Its negative association with children's publication output disappeared after adjustment, whereas its negative association with children's mean FWCI remained significant (Table S4). Children of national academy members also had lower Field similarity with their parents than children of non-members. However, Field similarity was not significantly associated with children's odds of national academy membership or prizewinning status after adjustment (Table S6). Thus, greater research differentiation was associated with higher field-normalized citation impact, but not with publication output or elite recognition, cautioning against interpreting divergence as a general pathway to academic achievement.

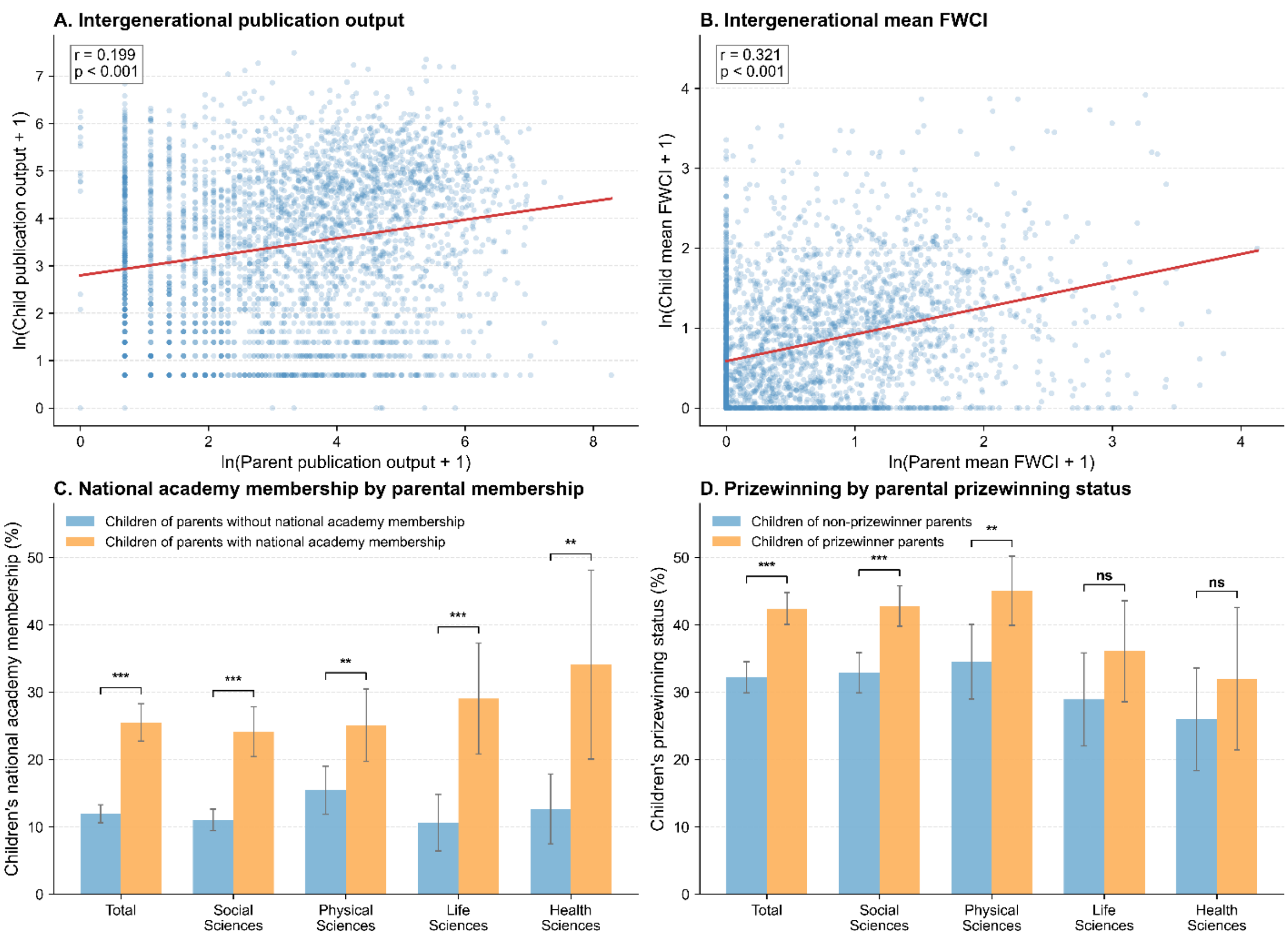

**Figure 3.** Intergenerational continuity in academic achievement and recognition. (A) Association between parents' and children's publication output. (B) Association between parents' and children's mean FWCI. (C) Children's national academy membership by parental membership status. (D) Children's recorded prizewinning status by parental prizewinning status. Error bars in panels C and D represent 95% confidence intervals.

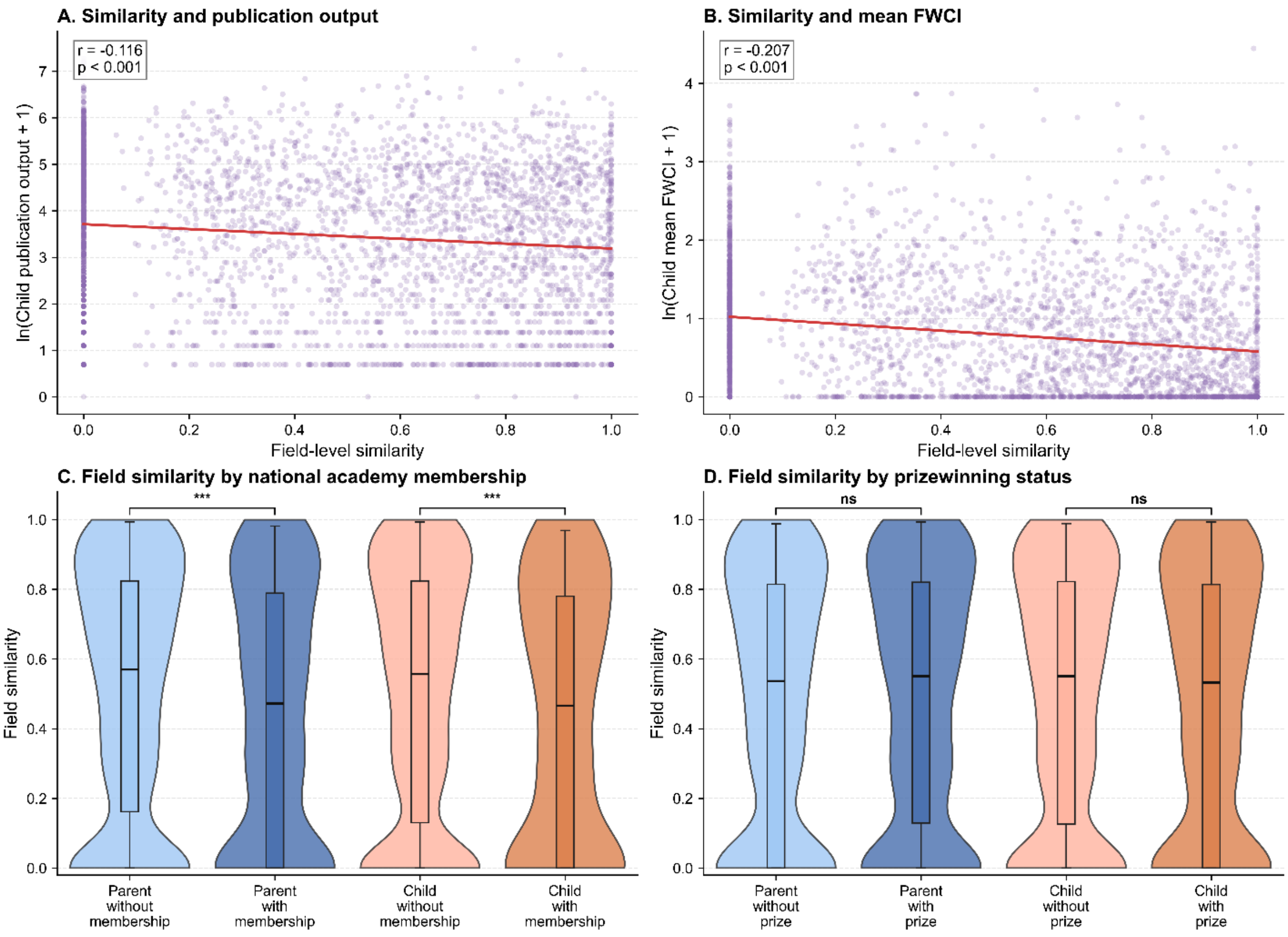

**Figure 4.** Field-level research similarity and children's academic outcomes. (A) Association between Field similarity and children's publication output. (B) Association between Field similarity and children's mean FWCI. (C) Field similarity by parents' and children's national academy membership. (D) Field similarity by parents' and children's recorded prizewinning status.

# Discussion

This study shows that parent–child pairs had substantial heterogeneity in Field-level research similarity, with both intergenerational continuity and differentiation commonly occurring. This coexistence suggests that academic families do not transmit research orientations in a uniform manner. Parents' academic resources may make related research areas more accessible to the next generation by providing disciplinary knowledge, professional networks, collaborators, and familiarity with particular academic institutions (11, 13, 20, 22). However, access to these resources does not always lead scholars' children to reproduce their parents' research paths. The same advantages may also provide them with the security and opportunities needed to pursue distinct research directions.

This interpretation helps reconcile cumulative advantage with niche partitioning. Rather than treating the two perspectives as mutually exclusive explanations, they may describe different responses to inherited academic resources. Some scholars' children may build upon parental expertise and networks, whereas others may differentiate themselves to establish an independent scholarly identity and avoid being overshadowed by a prominent parent, particularly one who is a national academy member. Intergenerational continuity should therefore be understood as one possible use of inherited academic resources rather than an automatic consequence of growing up in an academic family.

Intergenerational research continuity varied across disciplinary contexts, with parent–child pairs in the Social Sciences exhibiting comparatively greater similarity in research orientation. One possible explanation lies in differences in disciplinary epistemic cultures (23, 24), including lower rates of collaboration in the social sciences (25). Knowledge production in the Social Sciences often places greater emphasis on individual interpretation, theoretical development, and independent authorship than on shared laboratories or experimental infrastructure (26, 27). Parents and their children may therefore remain within similar broad disciplinary environments without collaborating directly. In contrast, research in the Physical, Life, and Health Sciences more frequently depends on laboratories, specialized equipment, large datasets, and team-based production (28, 29). These organizational conditions may create more opportunities for collaboration and cross-citation between academically related family members. Nevertheless, the observed disciplinary differences may also reflect variation in publication and citation practices and should not be attributed solely to epistemic culture.

The relationship between research similarity and children's achievement was more complicated. Once relevant characteristics of both generations were considered, Field-level similarity was not independently associated with children's publication output but remained negatively associated with their field-normalized citation impact. Research differentiation may therefore be more relevant to the influence or recognition of children's work than to the volume of their scholarly production. Developing a distinct research direction may help scholars' children establish an independent intellectual identity and reduce the possibility that their contributions are perceived primarily as extensions of an established parental trajectory. This interpretation is consistent with evidence from mentorship networks that differentiation from mentors can accompany stronger career outcomes for mentees (18, 30).

The recognition results likewise provided mixed evidence for the association between research independence and academic recognition. Greater research differentiation occurred among national academy members but not among prizewinners, indicating that this pattern was not consistent across different forms of academic recognition. National academy membership and prizes may reflect different selection processes, career stages, and dimensions of scholarly reputation. Moreover, recognition recorded in Wikidata is necessarily incomplete and includes prizes of varying scholarly prestige, which may weaken observable relationships.

Overall, inherited academic advantage and intellectual independence appear to be compatible rather than opposing career pathways. Scholars' children may benefit from resources associated with their parents' academic positions without reproducing their parents' research orientations. However, causal direction remains uncertain: differentiation may promote independent recognition, successful scholars may have greater freedom to enter new research areas, or both may arise from unobserved disciplinary, institutional, and family circumstances.

Neither similarity nor differentiation should automatically be treated as evidence of dependence or merit. Similarity may reflect legitimate intellectual continuity, whereas differentiation may represent the development of an independent scholarly identity. Because academic status can shape the allocation of attention and credit (31), transparent assessment of individual contributions is particularly important when family ties and overlapping research interests complicate the attribution of scholarly credit.

This study has several limitations. First, Wikidata coverage is uneven and favors scholars with publicly documented biographies and established academic visibility. The sample therefore overrepresents prominent scholars, European countries, and the Social Sciences and may not reflect the experiences of the broader academic population. Moreover, the analysis includes only parent–child pairs for whom both scholars could be matched to OpenAlex author profiles. Future research could combine bibliometric data with national researcher registries, institutional records, or funding databases to construct more representative samples of academic families.

Second, both Wikidata and OpenAlex are subject to missing information and classification errors. Although the entity-matching procedure was evaluated using a gold-standard dataset, automated disambiguation cannot eliminate all false matches or missed records. OpenAlex research classifications may also represent historical scholars and scholars from less extensively indexed disciplines less accurately. In addition, academic awards and academy membership were identified from available Wikidata records and therefore depend on the completeness and consistency of these data. Future studies could combine automated matching with stratified manual validation and additional data sources.

Third, the study identifies associations rather than causal effects. Research similarity cannot establish that parents directly transmitted particular research interests to their children, and the negative association between similarity and children's mean field-normalized citation impact does not demonstrate that research divergence produces greater citation impact. Unobserved factors, including education, institutional affiliation, and collaboration networks, may influence both research direction and achievement.

Finally, the OpenAlex profiles used here summarize scholars' research across their recorded careers, potentially concealing changes within individual trajectories. Intergenerational similarity may evolve from early collaboration or parental proximity to greater research independence later in the career of a scholar's child. Longitudinal research using publication-level time series could examine when such convergence and divergence occur and how they relate to collaboration and academic recognition across career stages.

In general, the findings support a cautious interpretation: academic families transmit resources, recognition, and access to research environments, but this transmission does not require children to reproduce parental research paths. Similarity and differentiation should therefore be interpreted as alternative ways in which inherited academic resources can be used, rather than as simple indicators of dependence or merit.

# Materials and Methods

## Data sources and sample construction

We identified parent–child scholar pairs from Wikidata using recorded father and mother relationships. Both members of a pair were required to have scholar-related occupations within the Wikidata scientist hierarchy. For each identified scholar, we retrieved biographical and academic information, including name, gender, date of birth, country of citizenship, occupation, field of work, awards, organizational memberships, employers, and educational institutions. We also retrieved ORCID and OpenAlex identifiers for the scholars and Research Organization Registry (ROR) identifiers for their employers and educational institutions where available. These fields were used both to describe the sample and to support entity matching between Wikidata and OpenAlex. The full data construction workflow is provided in Fig. S1.

Because Wikidata does not provide complete publication records, scholar entities were linked to OpenAlex author profiles. For records containing ORCID or OpenAlex identifiers, we retrieved the corresponding OpenAlex author profile directly. For records without direct identifiers, we searched OpenAlex using the scholar's English name and retained the first five author profiles returned by OpenAlex's default search-relevance ranking. Gemini 2.5 Flash-Lite was then used to compare Wikidata information with OpenAlex candidate profiles, including names, descriptions, occupations, fields, education, affiliations, ROR identifiers, topics, concepts, and historical affiliations (see Supplementary Methods, "LLM prompt for candidate assessment"). Each candidate received a matching probability and confidence score. We retained candidates for which both scores were at least 0.70, removed implausible matches based on age at first publication, and resolved remaining multiple matches using the product of probability and confidence, followed by citation count, publication count, and original search rank as tie-breakers.

The threshold was evaluated using a gold-standard dataset of Wikidata scholars with recorded ORCID or OpenAlex identifiers. These identifiers were withheld during matching and used only to define reference OpenAlex profiles. The 0.70 and 0.80 thresholds produced the joint-highest F1-score. We selected 0.70 because it retained more matches; its estimated precision was 83.46% (Table S2). The final analytical sample contained 3,229 parent–child scholar pairs, 5,877 unique scholars linked to OpenAlex author profiles, and 488,659 publications associated with these scholars.

## Research similarity and matched baseline

OpenAlex organizes research classifications hierarchically into Domains, Fields, Subfields, and Topics. Its author-level disciplinary profiles aggregate the subject classifications of an author's publications and assign a weight to each category at each level. For each parent–child pair and disciplinary level, we aligned the parent's and child's category-weight vectors and normalized them into probability distributions. Let $P = (p_1, \ldots, p_n)$ and $Q = (q_1, \ldots, q_n)$ denote the normalized disciplinary distributions of the parent and child, and let $M = (P + Q)/2$. Kullback-Leibler divergence (32) was calculated as:

$$\mathrm{KLD}(P \parallel M) = \sum_{i=1}^{n} p_i log_2(\frac{p_i}{m_i})$$

where $p_i$ and $m_i$ are the probabilities assigned to category $i$ in $P$ and $M$, respectively, and terms with $p_i = 0$ are defined as zero. $\mathrm{KLD}(Q \parallel M)$ is defined analogously by replacing $p_i$ with $q_i$. The Jensen–Shannon divergence (JSD) between the parent and child distributions is then calculated as:

$$\mathrm{JSD}(P \parallel Q) = \frac{1}{2}\,\mathrm{KLD}(P \parallel M) + \frac{1}{2}\,\mathrm{KLD}(Q \parallel M)$$

We converted divergence into similarity as Similarity $(P, Q) = 1 - \mathrm{JSD}(P \parallel Q)$. The resulting score ranges from 0 to 1, where 1 indicates identical disciplinary distributions and 0 indicates no overlap. The OpenAlex classification used here comprised 4 Domains, 26 Fields, 252 Subfields, and 4,516 Topics. The main analyses focus on Field-level similarity because similarity scores at the Topic and Subfield levels showed strong floor effects, whereas Domain-level scores showed ceiling effects. Results across all four OpenAlex levels are reported in Fig. S8.

To assess whether observed parent–child similarity exceeded what would be expected among comparable scholars from the same historical period, we constructed a matched random baseline. For each actual parent–child pair, the parent was retained and the child was replaced by a randomly selected comparison scholar from the children pool whose mean publication year fell in the same decade as that of the actual child. Other actual children of the same parent were excluded from the candidate pool. This random-matching procedure was repeated 1,000 times for each pair. Two stricter baselines additionally matched on children's gender and on children's gender and continent, respectively. These baselines are descriptive benchmarks rather than causal control groups because all pseudo-children were drawn from the same visibility-biased Wikidata dataset.

## Outcomes and statistical analysis

Academic interaction was measured using three binary outcomes: whether the parent and children had co-authored at least one publication, whether the parent had cited the children's work, and whether the children had cited the parent's work. Academic performance was measured using publication count and mean field-weighted citation impact (FWCI), both transformed using ln(1+x) in the correlation and regression analyses. Academic recognition was measured using national academy membership and recorded prizewinning status from Wikidata. Prizewinning status includes awards recorded in Wikidata and therefore varies in prestige and coverage.

Bivariate comparisons used non-parametric tests, chi-squared tests, McNemar's test for paired citation directions, and Pearson correlations as appropriate. Multivariable interaction models used logistic regression and adjusted for both generations' Domains, gender, mean publication years, publication counts, career spans, and the number of years of overlap between the parents' and children's publication-career periods. Academic performance models used ordinary least squares for log-transformed children's publication count and log-transformed children's mean FWCI, adjusting for corresponding parental performance, Field similarity, period, career span, gender, Domain, and parental recognition. Academic recognition models used logistic regression for children's national academy membership and prizewinning status, adjusting for the corresponding parental recognition status, the other parental recognition status, parental publication count, parental mean FWCI, period, career span, gender, and Domain. Standard errors were clustered by both parent and child identifiers. Adjusted model estimates and robustness checks are reported in the Supplementary Materials (Tables S3–S6 and S9).

# Acknowledgments

Zhichao Fang is financially supported by the National Natural Science Foundation of China (No. L2524083; No. 72304274). Er-Te Zheng is financially supported by the GTA scholarship from the School of Information, Journalism and Communication of the University of Sheffield.

## Ethics statement

This study was reviewed and approved by the University of Sheffield Research Ethics Committee (reference number: 075138). It used publicly available records from Wikidata and OpenAlex and involved no direct interaction with human participants.

## Use of generative AI

OpenAI GPT-5.5 was used to assist with language editing of the manuscript and to help write data-analysis code. The authors reviewed and revised all AI-assisted text, inspected and tested all AI-assisted code, and verified the reported analyses and results. The authors take full responsibility for the content of the manuscript, the analytical code, and the reported findings.

# Supporting Information for

## Should children follow their parents' research paths? Intergenerational research continuity and divergence in academic families

Er-Te Zheng[1], Xiaorui Jiang[1], Zhichao Fang[2,3*], Mike Thelwall[1*]

[1] School of Information, Journalism and Communication, University of Sheffield, Sheffield, UK.

[2] School of Information Resource Management, Renmin University of China, Beijing, China.

[3] Centre for Science and Technology Studies (CWTS), Leiden University, Leiden, The Netherlands.

Corresponding authors: Zhichao Fang, E-mail: z.fang@cwts.leidenuniv.nl; Mike Thelwall, E-mail: m.a.thelwall@sheffield.ac.uk

# Overview

The Supplementary Information contains supplementary methods information and two sets of additional analyses. The Supplementary Methods section provides further details on dataset construction and entity matching. Section S1 uses a matched random baseline to assess whether the observed parent–child research similarity exceeds what would be expected among scholars from comparable historical career periods. Section S2 examines whether the main results are sensitive to the probability and confidence thresholds used to link Wikidata scholars to OpenAlex author profiles, by replicating key analyses at thresholds of 0.60 and 0.80 in addition to the main threshold of 0.70.

# Supplementary Methods

### Data construction and entity matching

Fig. S1 shows the complete workflow for dataset construction, entity matching, and research similarity analysis. The resulting dataset comprises 3,229 parent–child pairs, with descriptive statistics for both generations presented in Table S1. Parents were predominantly male and active in the mid-twentieth century, while children show a higher proportion of female scholars. Both generations are geographically concentrated in Europe and North America, and predominantly active in the Social Sciences and Physical Sciences.

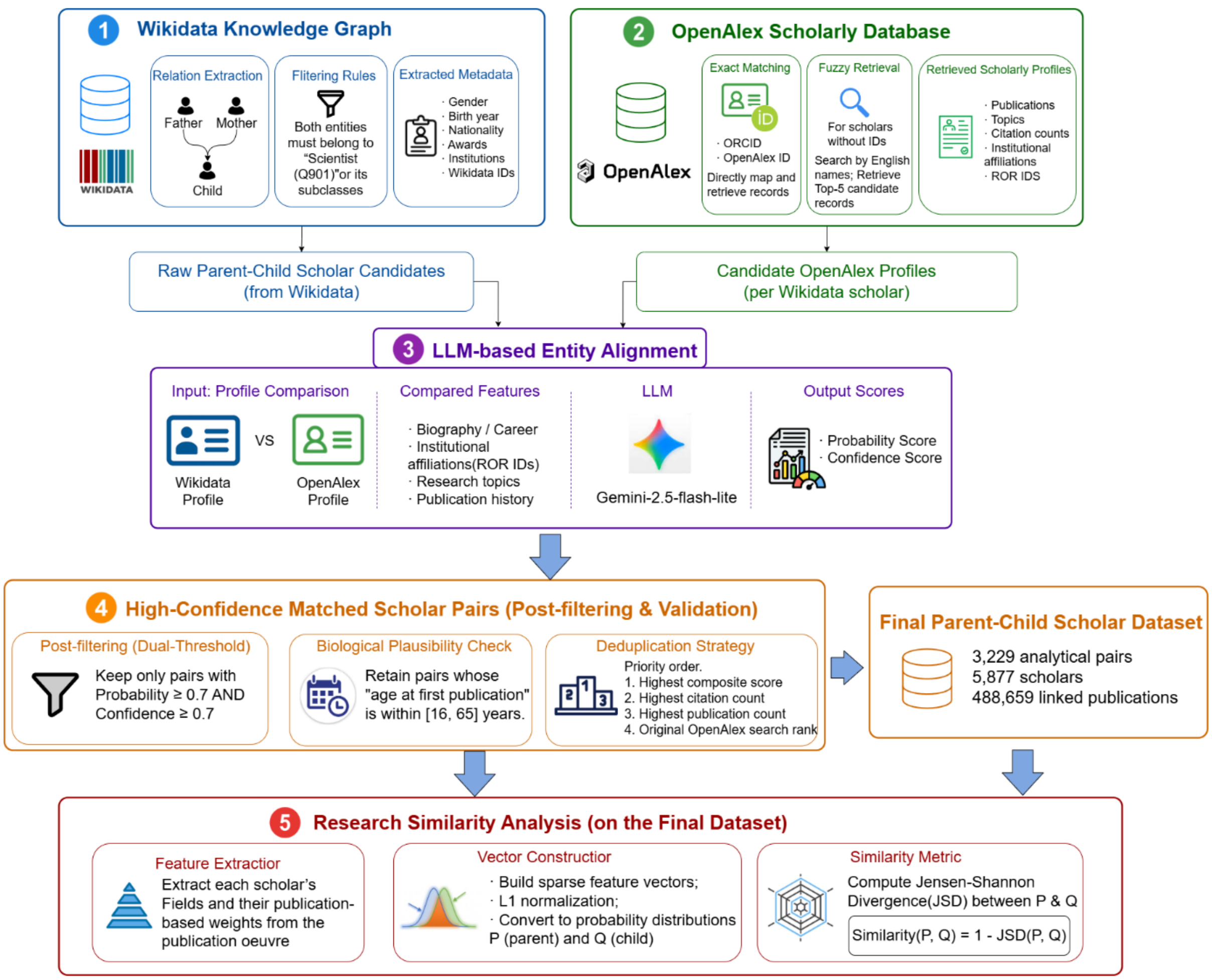

**Figure S1.** Workflow for constructing the parent–child scholar dataset, linking Wikidata entities to OpenAlex author profiles, and measuring intergenerational research similarity.

**Table S1. Descriptive statistics of the parent–child dataset.**

| Category | Sub-category | Parent | Child |
|---|---|---|---|
| Gender | Male | 2759 (85.4%) | 2285 (70.8%) |
| | Female | 468 (14.5%) | 932 (28.9%) |
| | Other | 1 (0.0%) | 3 (0.1%) |
| | Missing / Unknown | 1 (0.0%) | 9 (0.3%) |
| Has Award* | | 1674 (51.8%) | 1211 (37.5%) |
| National Academy membership** | | 945 (29.3%) | 514 (15.9%) |
| Average Publication Year | Mean | 1969.1 | 1990.2 |
| | Missing / Unknown | 62 (1.92%) | 30 (0.93%) |
| Continent*** | Europe | 2222 (68.8%) | 2027 (62.8%) |
| | North America | 541 (16.8%) | 567 (17.6%) |
| | Asia | 207 (6.4%) | 179 (5.5%) |
| | South America | 44 (1.4%) | 43 (1.3%) |
| | Oceania | 40 (1.2%) | 32 (1.0%) |
| | Africa | 18 (0.6%) | 12 (0.4%) |
| | Transcontinental | 13 (0.4%) | 2 (0.1%) |
| | Missing / Unknown | 276 (8.5%) | 468 (14.5%) |
| Domain | Social Sciences | 2007 (62.2%) | 1928 (59.7%) |
| | Physical Sciences | 646 (20.0%) | 725 (22.5%) |
| | Life Sciences | 324 (10.0%) | 374 (11.6%) |
| | Health Sciences | 202 (6.3%) | 179 (5.5%) |
| | Missing / Unknown | 50 (1.5%) | 23 (0.7%) |

*Any award recorded in Wikidata.

**Membership in a recognized national or royal academy, royal society, or Academia Europaea, as recorded in Wikidata.

***Scholars associated with multiple countries may be counted in more than one continent.

## Gold standard validation of entity matching

To validate the matching procedure, Wikidata scholars with recorded ORCID or OpenAlex identifiers were treated as the gold-standard sample. These identifiers were withheld during matching, and each scholar was rematched using the same candidate-retrieval, LLM-assessment, filtering, and deduplication procedure applied to scholars without direct identifiers. The selected OpenAlex ID was then compared with the reference ID in the gold-standard dataset. Precision, recall, and F1-scores were calculated across four joint probability and confidence thresholds (Table S2). Based on this validation, the main analyses used a joint threshold of 0.70 for both scores because it achieved the joint-highest F1-score while retaining more matches than the 0.80 threshold.

**Table S2. Gold-standard validation of entity matching across joint probability and confidence thresholds.**

| Joint threshold | Total gold samples | Retained predictions | Correct matches | Precision | Recall | F1-score |
|---|---|---|---|---|---|---|
| 0.6 | 1,407 | 1,210 | 1,009 | 0.8339 | 0.7171 | 0.7711 |
| 0.7 | 1,407 | 1,209 | 1,009 | 0.8346 | 0.7171 | 0.7714 |
| 0.8 | 1,407 | 1,196 | 1,004 | 0.8395 | 0.7136 | 0.7714 |
| 0.9 | 1,407 | 740 | 670 | 0.9054 | 0.4762 | 0.6241 |

### LLM prompt for candidate assessment

The following prompt was used to assess candidate OpenAlex author profiles in both the main matching procedure and the gold-standard validation. Terms enclosed in curly braces (e.g., {wiki_info_str}, {candidates_str}) are template placeholders replaced programmatically with the relevant scholar information before each query:

*"You are a rigorous academic data matching expert. Your task is to determine whether a scholar's profile from Wikidata and candidate profiles from OpenAlex refer to the exact same person.*

*### Matching Criteria (Crucial):*

*1. The Core Principle (Direction is King): Research Direction (Topic/Concepts vs. Description/Field) is the most critical factor. A match based solely on Institution with missing/vague Research Direction must NEVER receive a high probability score. However, if the Research Direction aligns perfectly, a high score can be given even if Institution data is missing.*

*2. Perfect Match (Very High Probability, High Confidence): Both Institution AND specific Research Direction are highly aligned. (Scores: Prob 0.90-0.99, Conf 0.90-0.99)*

*3. Scholar Mobility / Missing Institution (High Probability, High/Medium Confidence): Institutions differ or are missing on either side, BUT the specific sub-fields align perfectly (e.g., "archaeologist" matches "Archaeology Studies"). This strongly indicates the same person. (Scores: Prob 0.80-0.90, Conf 0.75-0.90)*

*4. Institution Match ONLY (Medium/Low Probability, Low Confidence): Institutions match, but Research Direction is completely missing or extremely vague on either side. It could be a namesake within a large university. Do not give a high score here. (Scores: Prob 0.30-0.50, Conf 0.30-0.50)*

*5. Same-Institution Name Collision (Low Probability, High Confidence): Institutions match, but research directions are completely different (e.g., Medicine vs. Literature). (Scores: Prob 0.05-0.20, Conf 0.80-0.95)*

*6. No Relation (Very Low Probability, High Confidence): Both institutions and research directions are completely different. (Scores: Prob 0.00-0.10, Conf 0.90-0.99)*

*7. Severe Information Missing (Low Probability, Low Confidence): Neither Institution nor Research Direction is sufficient to make any judgment. (Scores: Prob 0.10-0.30, Conf 0.10-0.30)*

*### Scoring Requirements:*

*- probability_score: 0.00 to 1.00, representing the likelihood that they are the same person.*

*- confidence_score: 0.00 to 1.00, representing how confident you are in your probability score based on the available evidence.*

*- Decimal Rule: Scores MUST be exact to two decimal places. Do NOT round to the nearest 0.05 or 0.10. Output granular and specific probabilities (e.g., 0.83, 0.12, 0.94).*

*### Input Data:*

*[Wikidata Scholar Information]*

*{wiki_info_str}*

*[OpenAlex Candidate Information]*

*{candidates_str}*

*### Output Format:*

*You must output strictly in valid JSON format. Do NOT include markdown formatting.*

*CRITICAL: The keys of your JSON MUST be the EXACT strings provided after “OpenAlex Candidate ID:” in the input. Do not abbreviate them.*

*The format must be EXACTLY as follows:*

*{{*

*“https://openalex.org/A1234567890”: {{“probability_score”: 0.83, “confidence_score”: 0.91}},*

*“https://openalex.org/A0987654321”: {{“probability_score”: 0.12, “confidence_score”: 0.88}}*

*}}”*

## Supplementary regression tables

**Table S3. Associations between Field similarity and parent–child academic interaction.**

| Focal predictor | Outcome | OR | 95% CI | N |
|---|---|---|---|---|
| Field similarity (+0.1) | Co-authorship | 1.38*** | [1.32, 1.45] | 3,133 |
| Field similarity (+0.1) | Parent cites child | 1.39*** | [1.31, 1.48] | 3,133 |
| Field similarity (+0.1) | Child cites parent | 1.39*** | [1.33, 1.47] | 3,133 |

*Notes: Logistic regression odds ratios are reported for a 0.1 increase in Field-level similarity. Models adjust for both generations' Domains, gender, mean publication years, publication counts, career spans, and duration of overlap between the parents' and children's publication-career intervals. Standard errors are two-way clustered by parent and child. *** $p<0.001$.*

**Table S4. Associations between parental performance, Field similarity, and children's academic performance.**

| **Focal predictor** | **Outcome** | **Coefficient** | **SE** | **N** |
|---|---|---|---|---|
| Parent publication output | Child publication output | 0.135*** | 0.018 | 3,148 |
| Parent mean FWCI | Child mean FWCI | 0.237*** | 0.025 | 3,050 |
| Field similarity | Child publication output | -0.042 | 0.061 | 3,133 |
| Field similarity | Child mean FWCI | -0.177*** | 0.038 | 3,050 |

*Notes: OLS coefficients are reported. Publication counts and mean FWCI are transformed using ln(1+x). Models adjust for publication period, career span, gender, Domain, and parental recognition. Standard errors are two-way clustered by parent and child. *** p<0.001.*

**Table S5. Associations between parents' and children's academic recognition.**

| Focal predictor | Outcome | OR | 95% CI | N |
|---|---|---|---|---|
| Parent national academy membership | Child national academy membership | 2.04*** | [1.59, 2.61] | 3,080 |
| Parent prizewinning status | Child national academy membership | 0.94 | [0.75, 1.18] | 3,080 |
| Parent prizewinning status | Child prizewinning status | 1.46*** | [1.24, 1.73] | 3,080 |
| Parent national academy membership | Child prizewinning status | 0.88 | [0.73, 1.06] | 3,080 |

*Notes: Logistic regression odds ratios are reported. Models adjust for publication period, career span, gender, Domain, parental publication count, and parental mean FWCI. Standard errors are two-way clustered by parent and child. *** $p<0.001$.*

**Table S6. Academic recognition and intergenerational Field similarity.**

| Model | Focal predictor | Outcome | Estimate | 95% CI | N |
|---|---|---|---|---|---|
| OLS | Parent national academy membership | Field similarity | -0.035* | [-0.063, -0.007] | 3,133 |
| OLS | Parent prizewinning status | Field similarity | 0.004 | [-0.021, 0.028] | 3,133 |
| Logit | Field similarity | Child national academy membership | 0.824 | [0.599, 1.135] | 3,075 |
| Logit | Field similarity | Child prizewinning status | 0.870 | [0.687, 1.103] | 3,075 |

*Notes: Estimates are OLS coefficients for models of Field similarity and odds ratios for logistic models of children's recognition. OLS models adjust for gender, Domain, career span, and publication period. Logistic models adjust for parental recognition, parental publication count and mean FWCI, gender, Domain, career span, and publication period. Standard errors are two-way clustered by parent and child identifiers. * $p<0.05$.*

# Section S1. Matched random baseline analysis

## S1.1 Design

For each actual parent–child scholar pair, the parent was retained and the child was replaced by a pseudo-child randomly drawn from the pool of scholars' children whose mean publication decade matched that of the actual child. All other biological children of the same parent who appeared in the dataset, as well as the parent, were excluded from the candidate pool. This sampling procedure was repeated 1,000 times for each pair. The primary specification matched pseudo-children on mean publication decade; two additional robustness specifications further matched on children's gender and on children's gender and continent, respectively. Field-level research similarity was calculated for each pseudo-pair using the same procedure applied to the actual parent–child pairs. This generated a matched random baseline distribution representing the similarity expected if each parent were paired with a non-biological child from the same publication decade. The observed similarity of each actual parent–child pair was then compared with its pair-specific baseline distribution.

**Statistical tests.** Two complementary test statistics are reported for each baseline specification. First, a one-sided empirical p-value was calculated as the proportion of the 1,000 random-matching iterations in which the baseline median equalled or exceeded the actual-pair median, using the correction (count + 1)/(N + 1). Second, a one-sided Wilcoxon signed-rank test compares, across all pairs, each pair's observed Field similarity against its pair-specific baseline mean.

**Limitations.** The pseudo-child pool is drawn entirely from scholars who are themselves children of scholars in this dataset; it does not represent the general academic population. This internal baseline may be conservative because pseudo-children are drawn from the same visibility-biased data environment, but it should still be interpreted as a within-sample benchmark rather than a population control. The design controls for historical period but does not account for geographic or institutional clustering, socioeconomic sorting, or ability-based selection into particular fields.

## S1.2 Results

Table S7 presents, for each of the three matching specifications, the number of pairs included, the median similarity among actual parent–child pairs, the median of the baseline medians across 1,000 random-matching iterations, their difference, the empirical p-value, and the share of actual pairs whose similarity exceeded their pair-specific baseline mean.

**Table S7. Overall matched baseline results across three matching specifications.**

| **Baseline specification** | **N pairs** | **Observed median** | **Baseline median** | **Median difference** | **Empirical p-value** | **Share observed > baseline** |
|---|---|---|---|---|---|---|
| Decade only | 3,130 | 0.546 | 0.098 | 0.448 | 0.001 | 67.2% |
| Decade + gender | 3,114 | 0.546 | 0.111 | 0.435 | 0.001 | 66.9% |
| Decade + gender + continent | 3,000 | 0.548 | 0.158 | 0.390 | 0.001 | 66.1% |

*Notes: Median difference = observed median − mean matched baseline median. Empirical p-value: proportion of iterations in which the baseline median equalled or exceeded the observed median (with continuity correction). "Share observed> baseline" is the share of pairs whose observed similarity exceeded their pair-specific baseline mean.*

### S1.3 Figures

Supplementary Figs. S2–S3 display matched-baseline information not shown in the main text. Fig. S2 compares actual parent–child pairs with matched random comparison pairs across all four OpenAlex classification levels, while Fig. S3 summarizes robustness across stricter matching specifications.

To assess whether the matched-baseline pattern was specific to Field-level similarity or also visible across the OpenAlex hierarchy, we repeated the actual-versus-matched comparison at the Topic, Subfield, Field, and Domain levels. The random comparison pairs were drawn using the same publication-decade matching rule as the Field-level matched-baseline analysis.

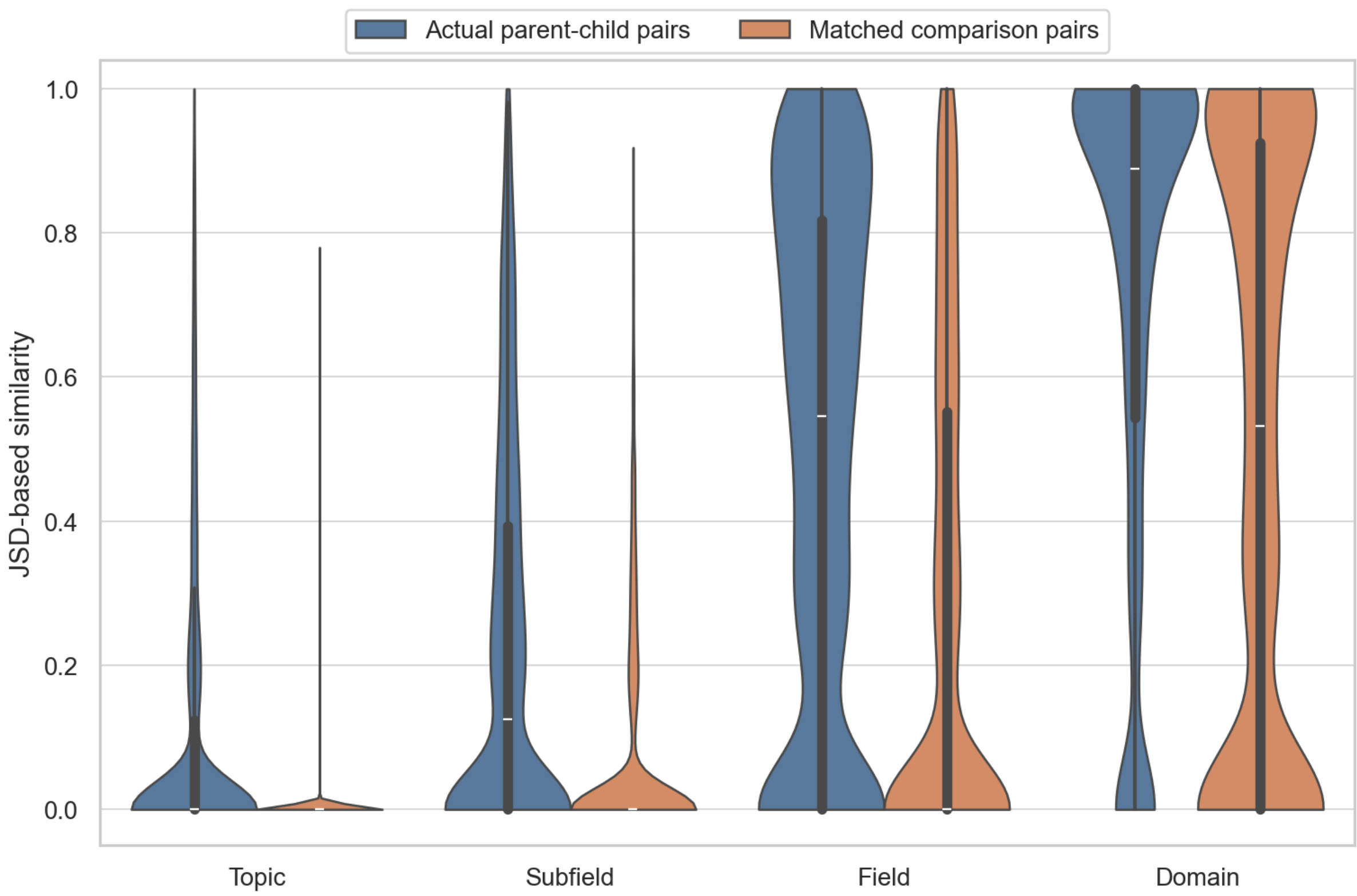

**Figure S2.** Distribution of research similarity at four OpenAlex classification levels for actual parent–child pairs and decade-matched random comparison pairs. For each pair, the parent was retained and the child was replaced by one randomly selected comparison scholar from the same publication decade as children, using the same candidate pools as the matched-baseline analysis.

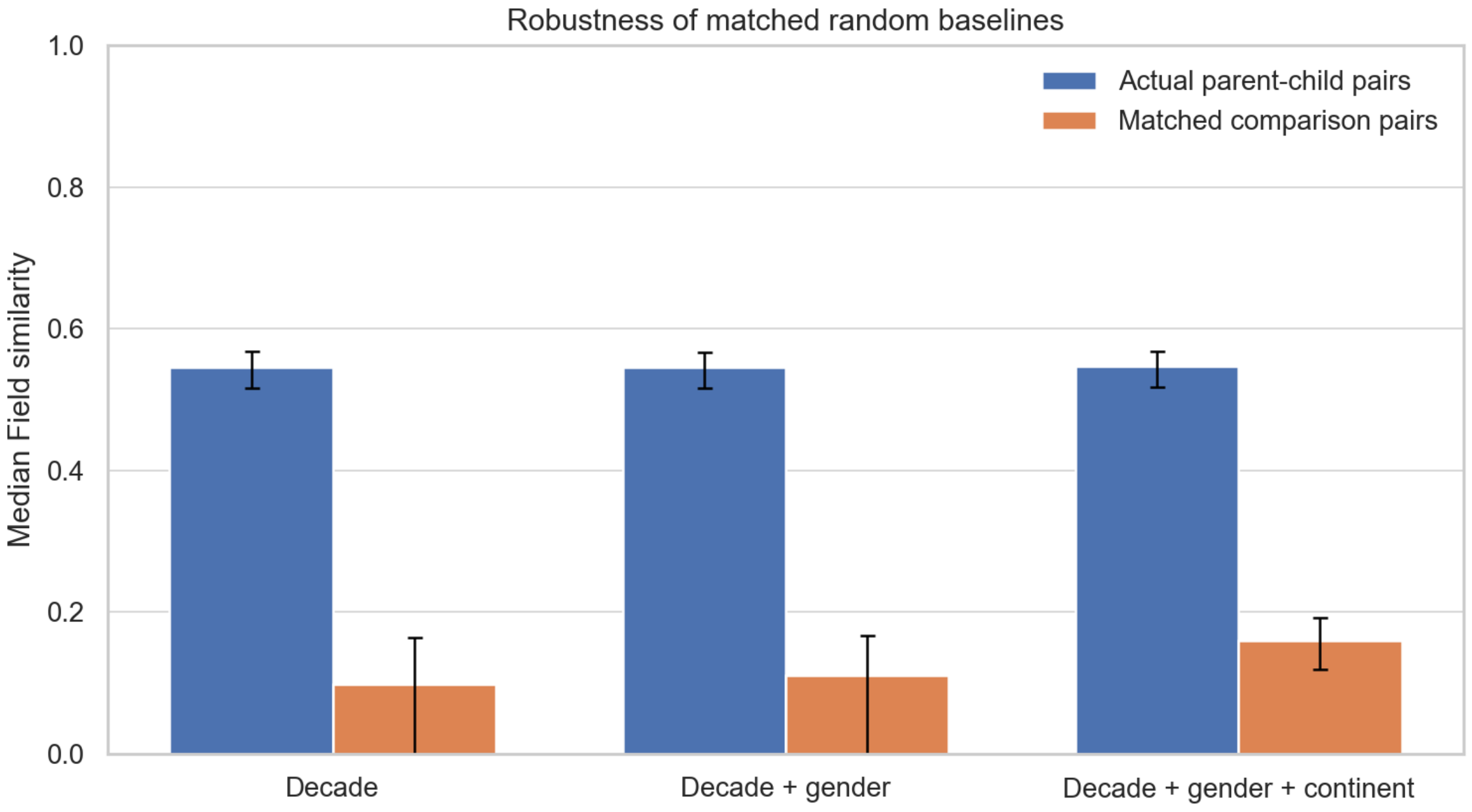

**Figure S3.** Robustness of the matched baseline across three matching specifications: decade only; decade and gender; decade, gender, and continent. Bars show median Field similarity with 95% confidence intervals.

# Section S2. Robustness checks across entity-matching thresholds

## S2.1 Design

Each scholar in the Wikidata parent–child list was linked to an OpenAlex author profile using two scores: a probability score (the likelihood that the candidate profile corresponds to the Wikidata entity) and a confidence score (the reliability of that assessment). The main analyses retain candidates whose probability and confidence scores both equal or exceed 0.70. To assess sensitivity to this choice, two alternative samples were constructed at thresholds of 0.60 and 0.80, applying the same biological-plausibility filter (first publication age 16–65 years), the same composite-score ranking, and the same one-to-one scholar-to-profile matching procedure. A preliminary 0.90 threshold retained only 576 parent–child pairs, compared with 3,229 in the main sample; sparse subgroups and unstable estimates led to its exclusion from the formal comparisons.

## S2.2 Sample sizes

Table S8 shows matched scholar counts and parent–child pair counts at each threshold, together with the number of pairs for which key variables are available. The main-threshold row (0.70) is highlighted in bold.

**Table S8. Analytical sample sizes across entity-matching thresholds.**

| Threshold | Matched scholars | Parent–child pairs | Pairs with Field similarity | Pairs with both mean publication years | Pairs with both mean FWCI |
|---|---|---|---|---|---|
| 0.60 | 5,958 | 3,278 | 3,207 | 3,145 | 3,047 |
| **0.70** | **5,877** | **3,229** | **3,161** | **3,148** | **3,050** |
| 0.80 | 5,198 | 2,842 | 2,790 | 2,783 | 2,700 |

*Notes: 'Pairs with Field similarity' counts pairs for which both parent and child Field profiles are non-missing. 'Pairs with both mean publication years' and 'Pairs with both mean FWCI' count pairs for which the respective metrics are available for both members.*

### S2.3 Robustness across alternative matching thresholds

Across all recorded analytical comparisons, the 0.60 and 0.80 threshold results retained the same directional classification and inferential outcome as the 0.70 main-analysis results. The results remained fully consistent across the five analytical domains examined: Field-similarity patterns, academic interaction, academic performance, academic recognition, and distributional comparisons. No comparison changed direction or crossed a significance threshold under the alternative matching specifications.

### S2.4 Core estimates across matching thresholds

Table S9 presents the focal estimates for each set of analyses at all three thresholds. Panel A reports Pearson correlations between children's mean publication year and Field similarity, assessing whether intergenerational research similarity varies across historical cohorts. Panel B reports differences in academic interaction between higher and lower Field similarity groups, including co-authorship, parent-to-child citation, and child-to-parent citation. Panel C reports OLS regression coefficients from models examining the associations of parental outcomes and Field similarity with children's academic performance outcomes (log-transformed publication count and mean FWCI), including full controls (gender, Domain, mean publication year, career span) and two-way clustered standard errors. Panel D reports recognition analyses in two sequential steps. Rows 1–2 report odds ratios (ORs) from logit regressions estimating the association between parental recognition status and children's recognition status. Rows 3–4 (Step 1) report OLS regression coefficients estimating the association between parental recognition status and Field similarity. Rows 5–6 (Step 2) report ORs from logit regressions estimating the association between Field similarity and children's recognition status. All models use two-way clustered standard errors, clustering by parent and child Wikidata identifiers.

**Table S9. Key estimates across entity-matching thresholds.**

| | (1)<br>Threshold 0.60 | (2)<br>Threshold 0.70 | (3)<br>Threshold 0.80 |
|---|---|---|---|
| **Panel A. Temporal trend (Pearson's r)** | | | |
| Children's mean publication year and Field similarity | −0.042*<br>*(0.018)*<br>N = 3,169 | −0.042*<br>*(0.018)*<br>N = 3,147 | −0.046*<br>*(0.019)*<br>N = 2,778 |
| **Panel B. Academic interaction — higher vs. lower Field similarity (percentage-point difference)** | | | |
| Co-authorship rate | +10.96***<br>*(1.03)*<br>N = 3,207 | +11.13***<br>*(1.05)*<br>N = 3,161 | +12.79***<br>*(1.16)*<br>N = 2,790 |
| Parents cite children's work | +6.89***<br>*(0.87)*<br>N = 3,207 | +7.12***<br>*(0.88)*<br>N = 3,161 | +8.11***<br>*(0.98)*<br>N = 2,790 |
| Children cite parents' work | +8.95***<br>*(1.02)*<br>N = 3,207 | +9.21***<br>*(1.04)*<br>N = 3,161 | +10.74***<br>*(1.15)*<br>N = 2,790 |
| **Panel C. Academic performance (OLS, full controls, two-way clustered SE)** | | | |
| Model (1): Parent's log(1+publications) → child's publications (baseline) | 0.135***<br>*(0.018)*<br>N = 3,142 | 0.135***<br>*(0.018)*<br>N = 3,148 | 0.140***<br>*(0.019)*<br>N = 2,780 |
| Model (2): Field similarity → child's publications (controlling for parent's) | −0.047<br>*(0.061)*<br>N = 3,125 | −0.042<br>*(0.061)*<br>N = 3,133 | −0.054<br>*(0.063)*<br>N = 2,765 |
| Model (3): Parent's log(1+mean FWCI) → child's mean FWCI (baseline) | 0.237***<br>*(0.025)*<br>N = 3,044 | 0.237***<br>*(0.025)*<br>N = 3,050 | 0.248***<br>*(0.027)*<br>N = 2,697 |
| Model (4): Field similarity → child's mean FWCI (controlling for parent's) | −0.179***<br>*(0.038)*<br>N = 3,042 | −0.177***<br>*(0.038)*<br>N = 3,050 | −0.181***<br>*(0.040)*<br>N = 2,695 |

| | (1)<br>Threshold 0.60 | (2)<br>Threshold 0.70 | (3)<br>Threshold 0.80 |
|---|---|---|---|
| **Panel D. Academic recognition (full controls, two-way clustered SE)** | | | |
| Intergenerational transmission of national academy membership (logit OR; ref. = no parental membership) | 2.061***<br>*[1.609, 2.640]*<br>N = 3,074 | 2.042***<br>*[1.595, 2.614]*<br>N = 3,080 | 1.956***<br>*[1.513, 2.528]*<br>N = 2,723 |
| Intergenerational transmission of prizewinning status (logit OR; ref. = no parental prize) | 1.468***<br>*[1.245, 1.731]*<br>N = 3,074 | 1.462***<br>*[1.240, 1.724]*<br>N = 3,080 | 1.441***<br>*[1.211, 1.715]*<br>N = 2,723 |
| Step 1a — Parent academy membership → Field similarity (OLS β; ref. = no parental membership) | −0.035*<br>*(0.014)*<br>N = 3,125 | −0.035*<br>*(0.014)*<br>N = 3,133 | −0.035*<br>*(0.015)*<br>N = 2,765 |
| Step 1b — Parent prizewinning status → Field similarity (OLS β; ref. = no parental prize) | 0.004<br>*(0.012)*<br>N = 3,125 | 0.004<br>*(0.012)*<br>N = 3,133 | 0.003<br>*(0.013)*<br>N = 2,765 |
| Step 2a — Field similarity → child national academy membership (logit OR) | 0.815<br>*[0.591, 1.124]*<br>N = 3,067 | 0.824<br>*[0.599, 1.135]*<br>N = 3,075 | 0.743<br>*[0.530, 1.042]*<br>N = 2,717 |
| Step 2b — Field similarity → child prizewinning status (logit OR) | 0.872<br>*[0.688, 1.106]*<br>N = 3,067 | 0.870<br>*[0.687, 1.103]*<br>N = 3,075 | 0.802<br>*[0.627, 1.027]*<br>N = 2,717 |

*Notes: Panel A reports Pearson r values, with standard errors in parentheses. Panel B reports percentage-point differences between the higher and lower Field similarity groups, with standard errors of the differences in parentheses. Panel C entries reports OLS regression coefficients, with two-way clustered standard errors in parentheses. Panel D reports logit odds ratios with 95% confidence intervals in brackets for rows 1–2 and 5–6, and OLS β coefficients with standard errors in parentheses for rows 3–4. * $p < 0.05$; ** $p < 0.01$; *** $p < 0.001$.*

## S2.5 Robustness checks for figures across matching thresholds

Figs. S4–S9 report robustness checks for the main manuscript figures and additional distributional comparisons at thresholds 0.60, 0.70, and 0.80.

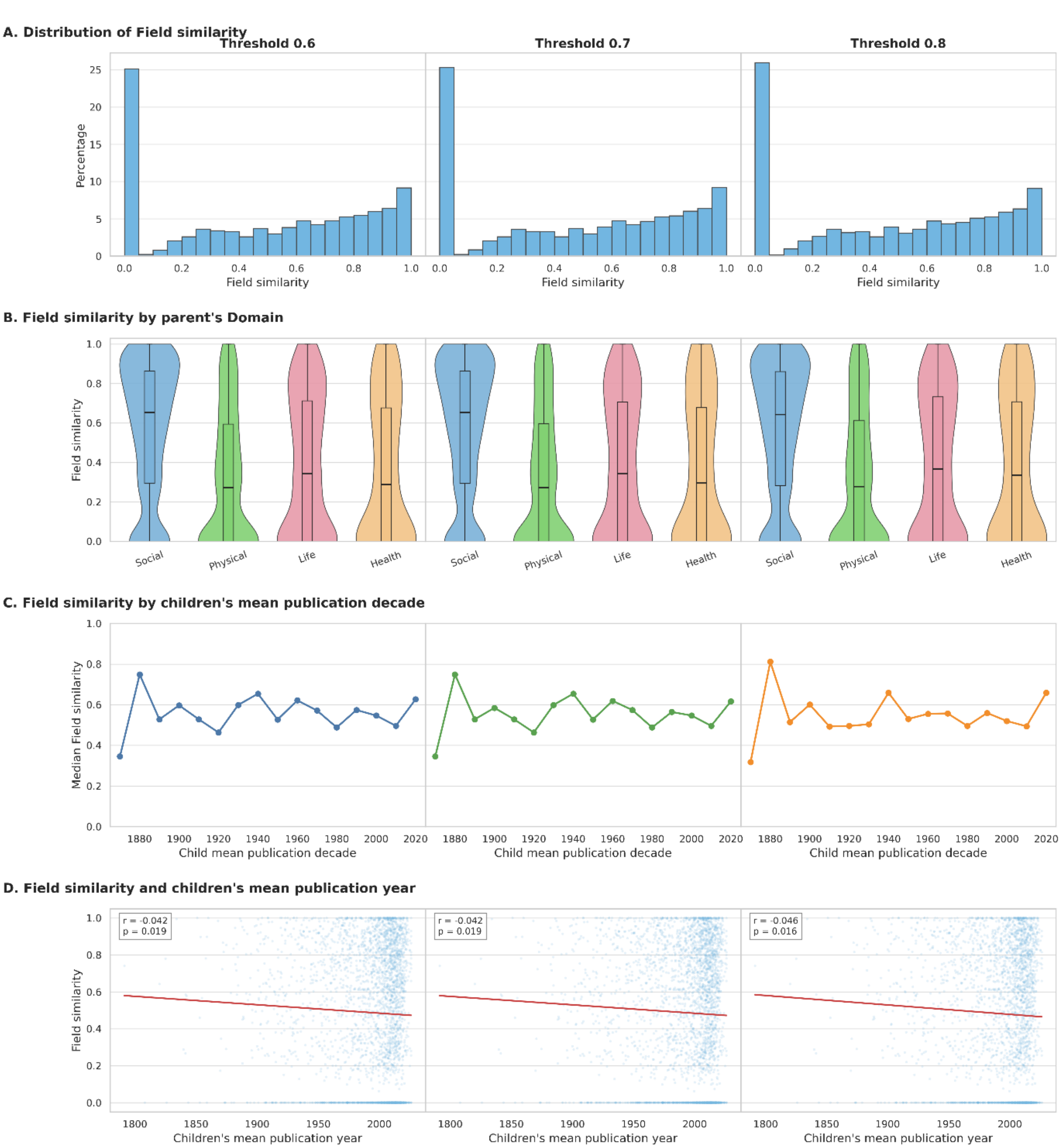

**Figure S4.** Robustness of Field-similarity distribution patterns corresponding to main Fig. 1. Each column corresponds to one entity-matching threshold. (A) Distribution of Field similarity. (B) Field similarity by parental Domain. (C) Median Field similarity by children’s mean publication decade. (D) Field similarity against children’s mean publication year.

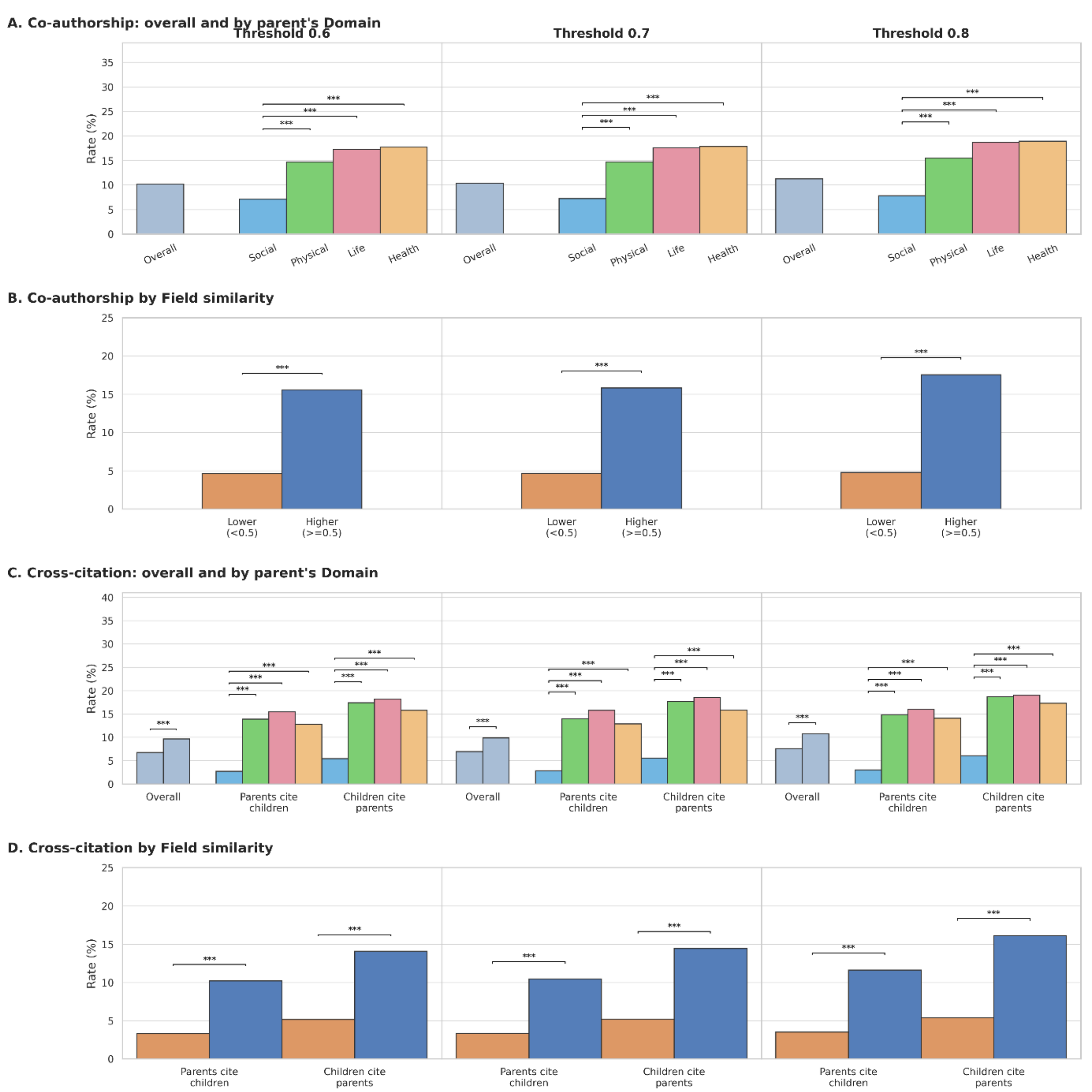

**Figure S5.** Robustness of co-authorship and cross-citation results corresponding to main Fig. 2. (A) Co-authorship rates overall and by parental Domain. (B) Co-authorship by Field-similarity group. (C) Cross-citation rates overall and by parental Domain. (D) Cross-citation by Field-similarity group.

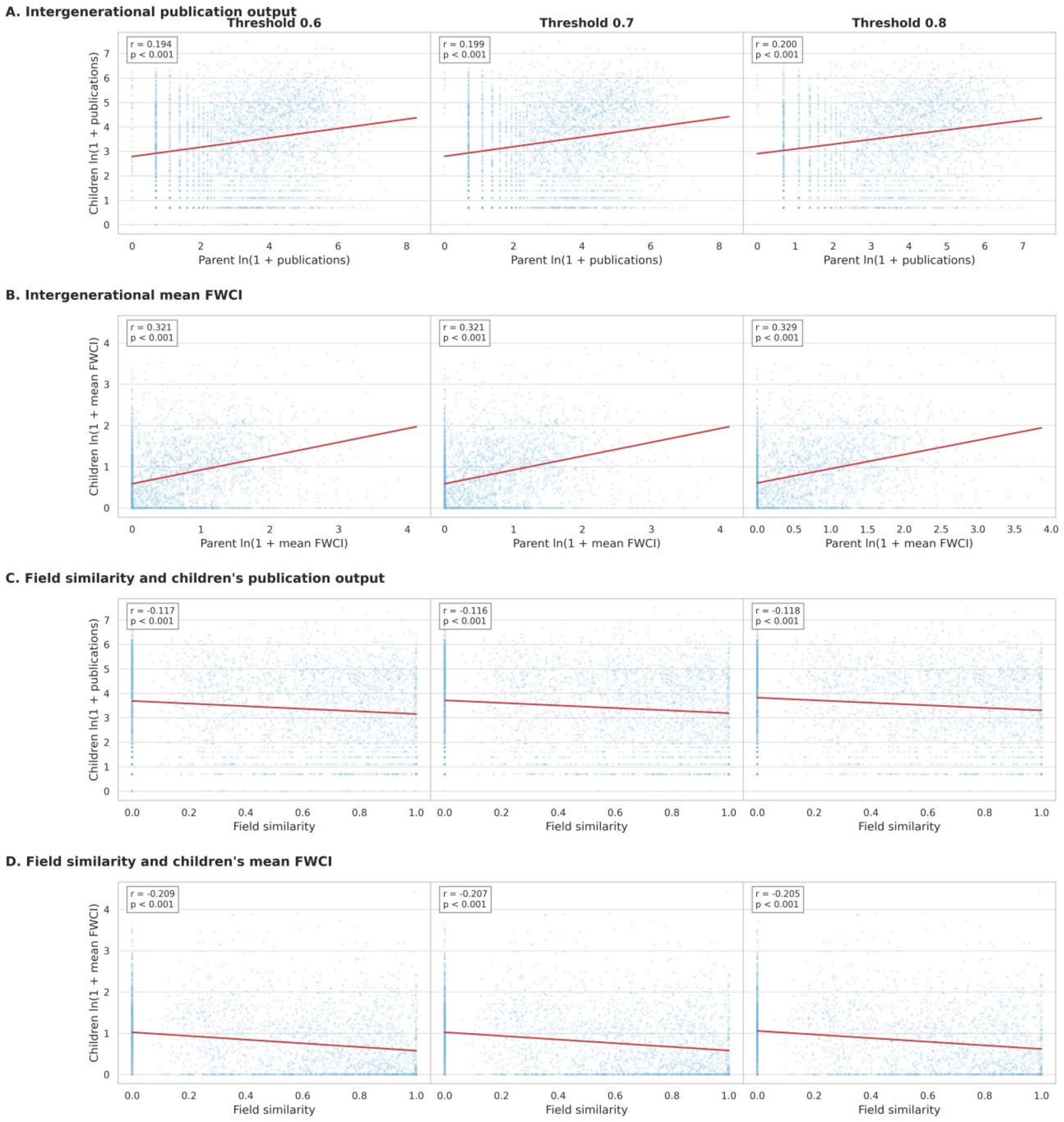

**Figure S6.** Robustness of academic-performance correlations corresponding to main Figs. 3A–B and 4A–B. (A) Intergenerational publication output. (B) Intergenerational mean FWCI. (C) Field similarity and children's publication output. (D) Field similarity and children's mean FWCI.

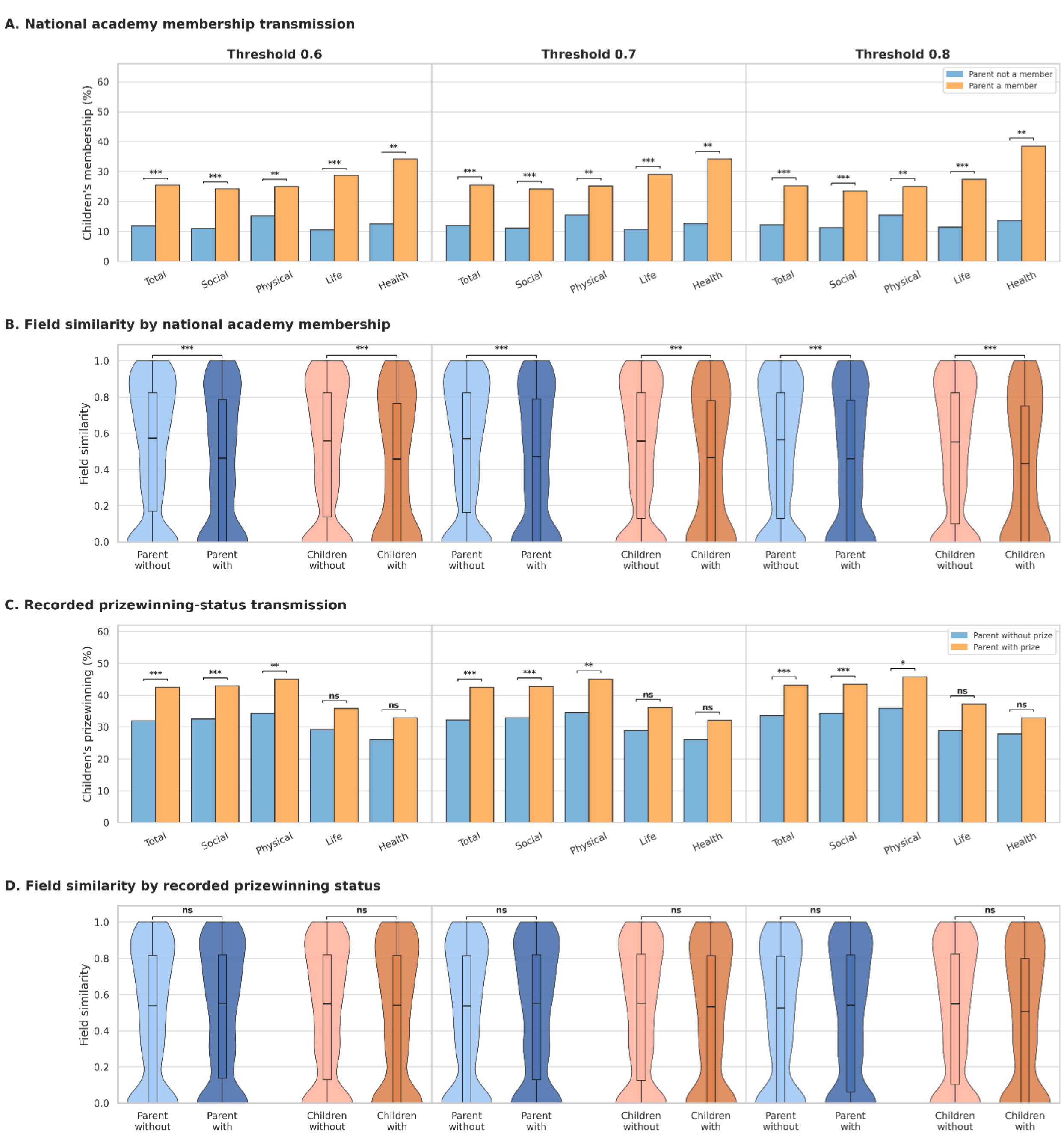

**Figure S7.** Robustness of academic-recognition results corresponding to main Figs. 3C–D and 4C–D. (A) National academy membership transmission. (B) Field similarity by national academy membership status. (C) Prizewinning-status transmission. (D) Field similarity by prizewinning status.

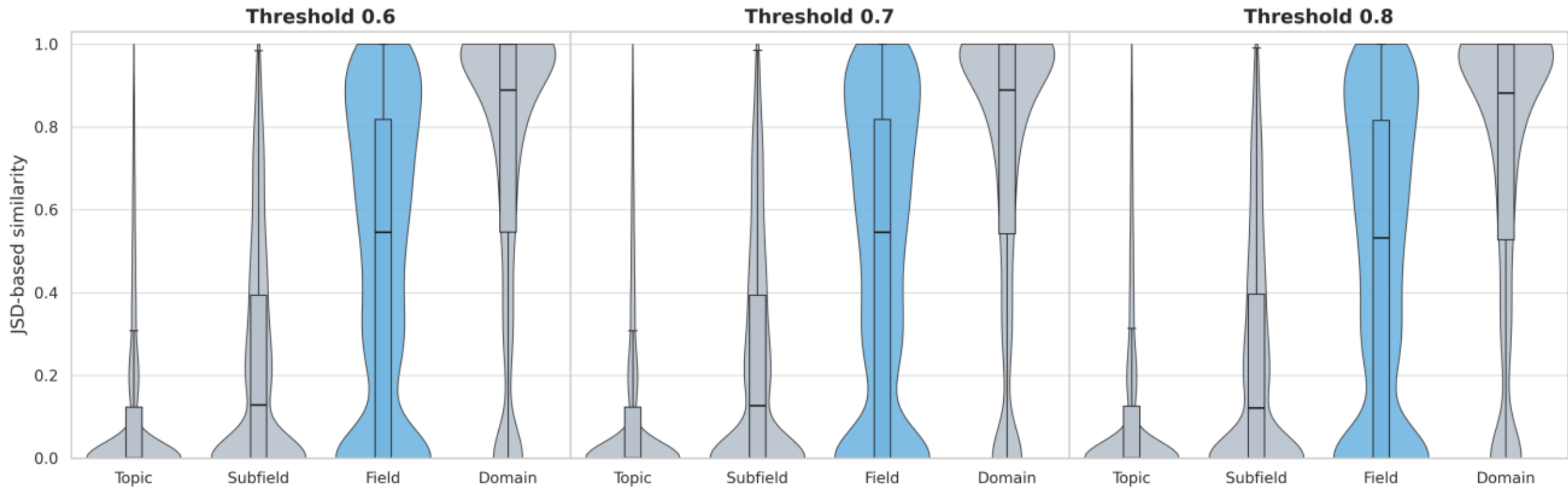

**Figure S8.** Robustness of research similarity distributions across OpenAlex classification levels (Topic, Subfield, Field, Domain).

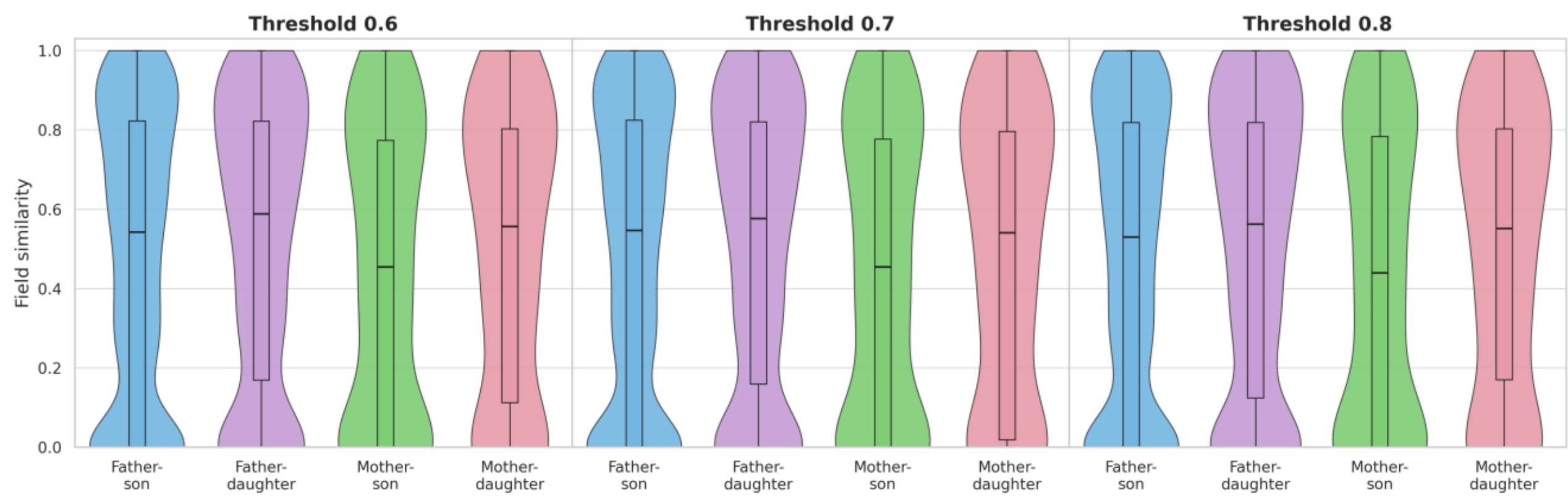

**Figure S9.** Robustness of Field similarity across parent–child gender combinations.